# Lasing from single, stationary, dye-doped glycerol/water microdroplets located on a superhydrophobic surface

A. Kiraz[a)], A. Sennaroglu[b)], S. Doğanay, M. A. Dündar, A. Kurt, and H. Kalaycıoğlu

*Department of Physics, Koç University, Rumelifeneri Yolu, 34450 Sariyer, Istanbul, Turkey*

A. L. Demirel

*Department of Chemistry, Koç University, Rumelifeneri Yolu, 34450 Sariyer, Istanbul, Turkey*

## Abstract

We report laser emission from single, stationary, Rhodamine B-doped glycerol/water microdroplets located on a superhydrophobic surface. In the experiments, a pulsed, frequency-doubled Nd:YAG laser operating at 532 nm was used as the excitation source. The microdroplets ranged in diameter from a few to 20 µm. Lasing was achieved in the red-shifted portion of the dye emission spectrum with threshold fluences as low as 750 J/cm$^2$. Photobleaching was observed when the microdroplets were pumped above threshold. In certain cases, multimode lasing was also observed and attributed to the simultaneous lasing of two modes belonging to different sets of whispering gallery modes.

[a)] akiraz@ku.edu.tr
[b)] asennar@ku.edu.tr

Optical microcavities confine high-quality optical resonances in small volumes [1,2]. They are therefore attractive in the development of ultralow-threshold lasers. Such light sources hold a great promise for applications in optical communications systems and fundamental studies in cavity quantum electrodynamics. Up to date, laser emission has been observed from various different optical microcavities based on semiconductors or polymers. Examples include microdisks [3], microspheres [4], micropillars [5], and photonic crystal defect microcavities [6].

Experiments performed more than 20 years ago demonstrated laser emission from liquid microdroplets flying in air [7,8,9]. In these experiments, a stream of dye-doped ethanol droplets was excited by a pulsed laser. Laser action was observed by analyzing the emission spectra collected from single microdroplets excited by single pump pulses. Electrodynamic levitation has often been used to analyze stationary droplets [10]. Recently lasing was also observed from levitated, Rhodamine 6G doped droplets [11]. To date, however, lasing has not been reported from stationary microdroplets standing on a substrate. One challenge in this case stems from the fact that the physical characteristics of the surface need to be carefully optimized in order to minimize the geometric deformation of the otherwise spherical droplet and to maintain a sufficiently high quality factor for the resonant modes.

Here, we report laser emission from single, stationary, dye-doped glycerol/water microdroplets located on a superhydrophobic surface. Due to the superhydrophobic nature of the surface, the spherical structure of the droplets is well maintained and a mode structure similar to that of an ideal spherical resonator can be obtained [12]. The technique used allows for the analysis of laser emission from a particular microdroplet over prolonged periods.

Superhydrophobic surfaces which are transparent to visible light were prepared by spin coating hydrophobically coated silica nanoparticles (Degussa AG, Aeroxide LE1) on cover glasses from 50 mg/ml dispersions in ethanol [13]. Average contact angle of 152.6° is measured for 2 mm diameter water droplets on these surfaces. A 90/10 water/glycerol solution doped with 225 µM Rhodamine B was sprayed onto a superhydrophobic surface using an ultrasonic nebulizer. Generated microdroplets were kept in contact with the ambient atmosphere. Once sprayed on the superhydrophobic surface, microdroplets quickly evaporated, and reached their equilibrium sizes with resulting diameters in the range from a few up to ~20 µm. Due to the uncertainty in the amount of evaporation during the formation of the microdroplets, it was not possible to exactly determine the concentration of the Rhodamine B dye. The estimated concentrations were in the 0.225-2.25 mM range.

Microdroplets were excited with a frequency-doubled, Q-switched Nd:YAG laser which produced green pulses at the wavelength of 532 nm. The pulse repetition rate and pulsewidth were 1 kHz and 100 ns, respectively. In the experiments, pump energies up to 40 µJ/pulse were available. Excitation was achieved by focusing the pump beam to a 14 µm diameter spot. A high numerical aperture microscope objective (60x, NA=1.4) was used in the inverted geometry both for excitation and collection of the fluorescence. The collected fluorescence was transmitted through a dichroic mirror, a 1.5x magnifier element, and was dispersed by a ½-meter monochromator. The spectra were then recorded with a silicon CCD camera. With the selected 300 grooves/mm grating and input slit width of 30 µm, a spectral resolution of 0.24 nm was achieved. A shutter was used to block the pump beam at all times except during the 50 ms exposure periods of the CCD camera. Pump power was adjusted using a variable-density filter.

Figures 1a-1c show the power dependent fluorescence spectra obtained from a 7.8 µm diameter microdroplet exhibiting laser emission. In Fig. 1a the intensity of the whispering gallery mode (WGM) at around 601 nm is comparable to the intensity of other WGMs and the background Rhodamine B emission. As shown in Fig.s 1b and 1c, this WGM dominates the emission spectrum at higher excitation powers. The observed nonlinear power dependence is an indication of laser emission. The lasing WGM is spectrally located in the red-shifted

portion of the Rhodamine B emission spectrum due to the overlap of the absorption and emission bands on the blue end of the dye emission spectrum, favoring lasing at red-shifted wavelengths. This is consistent with the previous demonstrations using microdroplets flying in air [7,8,14]. Due to the errors caused by the deviations from the ideal spherical geometry of the stationary microdroplets, no attempt was made to exactly identify the WGMs seen in Fig. 1. It is speculated that the lasing WGM has a mode order of 1 because of its high quality factor. Laser emission was observed to decay with photobleaching. Fig. 1d shows the emission spectrum recorded after a total irradiation time of 1.5 sec following the spectrum shown in Fig. 1c, under a constant excitation fluence of 4950 J/cm$^2$. As can be seen in Fig. 1d, the intensity of the WGM at around 601 nm becomes comparable to the intensities of the other WGMs as a result of photobleaching, and lasing is no longer observed in this case. In Fig. 1d, the free spectral range (FSR) of the lasing WGM is measured to be 13.5 nm. For an ideal glycerol microsphere (refractive index n=1.47, radius=7.8 µm), FSR is theoretically estimated to be 12.6 nm [8]. The difference between the theoretical estimate and the experimental result is explained by the nonspherical geometry of the microdroplet standing on a superhydrophobic surface [12].

We believe that, because the deformation of the spherical microdroplet is mainly in the direction perpendicular to the substrate, the observed high quality whispering gallery modes circulate parallel to the cover glass in the equatorial plane. In this plane, the cross-section of the microdroplets was still circular. Therefore the equatorial modes moving in both clockwise and anti-clockwise directions were still nearly degenarate. In support of this, the spectra reported in Fig.s 1 and 4 showed no frequency splitting within the spectral resolution of our spectrometer. This prevented the assignment of the direction (clockwise or anti-clockwise) of the lasing WGMs in the equatorial plane.

Power dependent intensities of the lasing WGM as well as two other WGMs and the background emission at 580 nm are plotted in Fig. 2. The emission from the WGMs at 587 nm and 592 nm, and the background emission at 580 nm show linear power dependences at low excitation powers. Above an excitation fluence of 1290 J/cm$^2$, they become saturated, without any clear indication of nonlinear power dependence. On the other hand, the lasing WGM exhibits nonlinear power dependence above a certain threshold excitation fluence. The threshold fluence was estimated from the crossing point of the linear least squares fits to the power-dependent emission data and was determined to be 750 J/cm$^2$. Saturation was observed in the lasing WGM near an excitation fluence of 3960 J/cm$^2$, where the slope efficiency decreased by more than a factor of 5.

Another indication of laser emission is the decrease in the spectral width of the lasing mode. This was observed for the specific lasing WGM shown in Figs. 1 and 2. The full-width-at-half-maximum (FWHM) of the lasing WGM was measured to decrease from 0.34 nm at the excitation fluence of 200 J/cm$^2$ to a resolution-limited width of 0.25 nm at excitation powers exceeding 1290 J/cm$^2$. In Fig. 3, we plot the dependence of the FWHM of another lasing WGM on the excitation fluence from a microdroplet with a diameter of 6.9 μm. In particular, the FWHM of this WGM is 0.54 nm at the relatively low excitation fluence of 990 J/cm$^2$. With the onset of lasing, the FWHM decreases down to 0.28 nm at the excitation fluence of 14840 J/cm$^2$.

We have also observed instances where multimode lasing occurred. This is shown in Fig. 4 where two WGMs oscillating at the wavelengths of 599 nm and 607 nm were seen to lase simultaneously in a 8.2 μm diameter microdroplet. As discussed above, uncertainties due to deviations from the ideal spherical geometry prevent the exact identification of the lasing WGMs. However, FSR measurements suggest that they belong to two different sets of WGMs having different mode orders or polarizations. The two sets of WGMs are indicated by the letters A and B in the figure. The measured FSR for the sets A and B are 12.68 nm and 13.09 nm, respectively. Similar to the microdroplet investigated in Fig. 1, the deviation of the observed FSRs from the theoretical estimate of 12.03 nm is explained by the nonspherical geometry of the microdroplet [12]. Estimations based on an ideal spherical microcavity suggest that the WGMs in set A and B can be identified as first order TE and TM WGMs respectively ($TE_{m,1}$ and $TM_{m,1}$). For an ideal spherical glycerol/water microcavity, the spectral separation between TE and TM WGMs with the same mode order and mode number will approach ~0.7·FSR (for radius >> wavelength) [15]. The spectral separation observed in Fig. 4 between the lasing WGMs (~0.65·FSR) fits well with this estimate. Due to their orthogonal polarizations, $TE_{m,1}$ and $TM_{m,1}$ WGMs will not compete for gain in an isotropic medium, and hence simultaneous lasing can be achieved.

As optical microcavities exhibiting laser emission, microdroplets standing on a superhydrophobic surface can be attractive alternatives to solid optical microcavities standing on substrates, e.g. microdisks, micropillars, photonic crystal defect microcavities. Microdroplets standing on a superhydrophobic surface do not pose any microfabrication challenges. They also bring together the advantage of easy deformability. This can lead to the realizations of tunable microcavity lasers by using evaporation/condensation [12], or electric field tuning [16]. In addition, lasing stationary microdroplets can be potentially used as an optical diagnostic tool in high-resolution surface characterization.

This work was supported by the Scientific and Technological Research Council of Turkey (Grant No. TÜBİTAK-105T500). The authors thank F. Menzel for providing the silica nanoparticles, and the Alexander von Humboldt Foundation for equipment donation. A. L. Demirel and A. Kiraz acknowledge the financial support of the Turkish Academy of Sciences in the framework of the Young Scientist Award program (Grants No. EA/TÜBA-GEBİP/2001-1-1 and A.K/TÜBA-GEBİP/2006-19).

Figure Captions

Figure 1: (a-c) Power dependent emission spectra obtained from a 7.8 µm diameter glycerol/water microdroplet. Excitation fluences are 200 J/cm$^2$, 940 J/cm$^2$, and 4950 J/cm$^2$ in (a), (b), and (c) respectively. Inset: Optical microscope image of the microdroplet, dashed circle shows the area illuminated by the excitation laser. (d) Emission spectrum recorded after a total irradiation time of 1.5 sec following the spectrum shown in (c), under a constant excitation fluence of 4950 J/cm$^2$.

Figure 2: Emission intensity vs. excitation fluence of a 7.8 µm diameter glycerol/water microdroplet for the lasing WGM (601 nm), two other WGMs (587 and 592 nm), and background emission at 580 nm.

Figure 3: (a) FWHM vs. excitation fluence for the lasing WGM of a 6.9 µm diameter glycerol/water microdroplet. (b) Emission intensity vs. excitation fluence for the lasing WGM (filled squares), and background emission at 580 nm (empty circles).

Figure 4: Emission spectrum obtained from an 8.2 µm diameter glycerol/water microdroplet exhibiting lasing at two WGMs at 599 and 607 nm. Excitation fluence is 6530 J/cm$^2$. Inset: Emission intensity vs. excitation fluence of the lasing WGMs, and background emission at 580 nm.

Figure 1

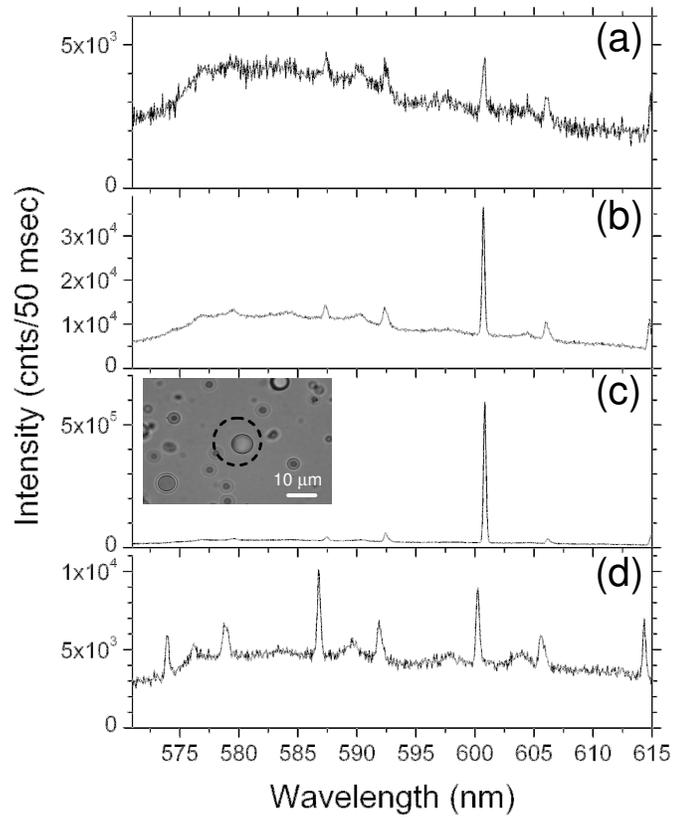

Figure 2

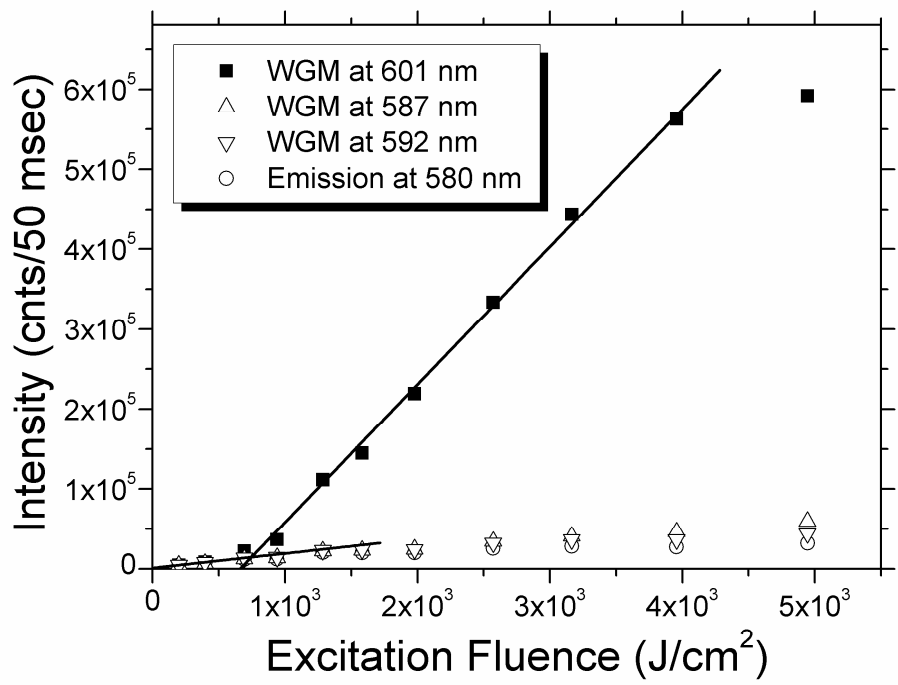

Figure 3

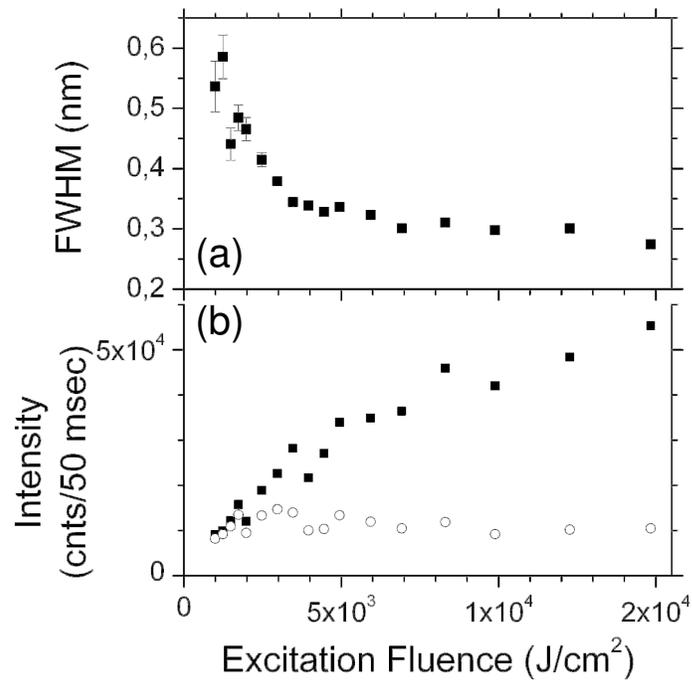

Figure 4

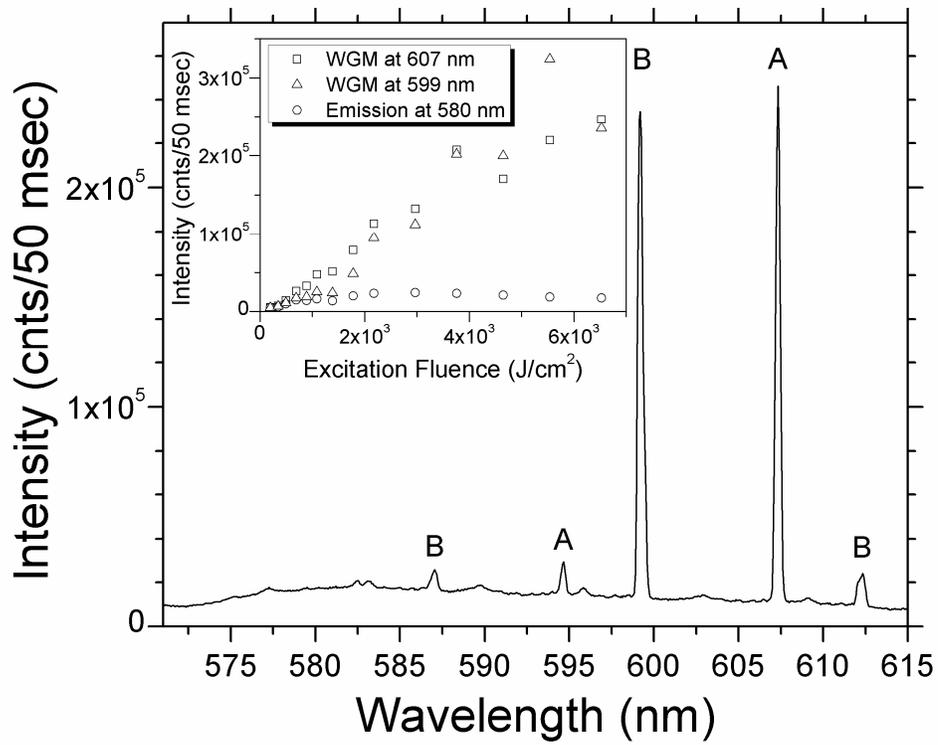